\documentclass[12pt]{iopart}
\usepackage{iopams}
\usepackage{graphicx}
\begin{document}

\title[Renormalization group description of continuous and the first
order phase transitions]{Renormalization group approach to unified
description of continuous and the first order phase transitions:
application to the Blume-Capel model}

\author{V. I. Tokar$^{1,2}$}

\address{$^1$Universit\'e de Strasbourg, CNRS, IPCMS, UMR 7504,
F-67000 Strasbourg, France}
\address{$^2$G. V. Kurdyumov Institute for Metal Physics of the 
N.A.S.\ of Ukraine, 36 Acad. Vernadsky Boulevard, UA-03142 Kyiv, Ukraine}
\begin{abstract}
The renormalization group (RG) equation in the self-consistent local
potential approximation (SC-LPA) suggested earlier for the description
of continuous phase transitions in lattice models of the Landau-Ginzburg
type has been applied to the solution of the spin-1 Blume-Capel model
on the simple cubic lattice.  The calculated transition temperatures
of both continuous and the first-order phase transitions (FOPTs) in
zero external field have been found to be in excellent agreement with
the best available estimates.  It has been argued that the SC-LPA RG
equation may give more accurate and complete description of the FOPTs
than those reported in alternative approaches.  It has been shown that
the SC-LPA RG equation can be cast in the form of the generalized Burgers'
equation (GBE). In this formulation of the RG the FOPTs have been shown to
assume the form of the shock-wave solutions of GBE in the inviscid limit.
Universality of the RG flow in the vicinity of the fixed point describing
FOPTs has been discussed.
\end{abstract} 

\maketitle
\section{Introduction}
From statistical-mechanical standpoint phase transitions are
nonperturbative phenomena because they are formally defined as
singularities in the dependence of the free energy (FE) on thermodynamic
parameters whereas the microscopic Hamiltonians at the transition
points are analytic.  Thus, the singularities cannot appear in a
finite order of a perturbative expansion and should be sought as the
points of divergence of the series.  In field-theoretic models of the
Landau-Ginzburg type high order perturbative expansions are notoriously
difficult \cite{free_params2RG2} so numerous alternative techniques have
been developed to describe phase transitions.

The most systematic approaches are based on the Monte Carlo (MC)
simulations \cite{binder} and the temperature series expansions (TSE)
\cite{n-vector-models,butera_blumecapel_2018}. However, reliable
quantitative results can be obtained within these techniques only
at the cost of large-scale numerical and/or analytic calculations
which still do not guarantee good accuracy because of the absence of
reliable convergence criteria. For example, in \cite{ferrenberg2018}
the MC simulations in the box containing more than a billion of lattice
sites performed during more than two thousand CPU core years did not
reproduce the critical exponent of the specific heat with acceptable
precision. Only one critical point was investigated in this study so
if one needs to deal with continuous lines of phase transitions the
computations may become prohibitively lengthy.

Such a situation is met in the Blume-Capel model (BCM)
\cite{blume_theory_1966,capel_possibility_1966} where phase
transitions comprise three lines of critical points, one line of
the first order phase transitions (FOPTs) and a tricritical point
(TCP) (see, e.g., figure 1 in \cite{butera_blumecapel_2018}).
For dealing with such cases many computationally undemanding
heuristic approaches have been developed for studies where
numerical accuracy is not a major priority (see, e.g., 
\cite{ngCVM,maxwell_construction,tokar_new_1997,scoza,latticeRG2010,%
tan_topologically_2011,parola_recent_2012,sc-lpa-rg} and
references therein).  Among them, the most prospective from the
standpoint of the unified description of phase transitions
of all kinds seems to be those based on the nonperturbative
renormalization group (NPRG) (see bibliography of in review papers
\cite{BAGNULS200191,berges_non-perturbative_2002}).  The main reason
for this is that such fundamental features of the continuous phase
transitions as the universality and the scaling laws are intrinsic to
the RG while even in the most rigorous alternative approaches their
validity is not guaranteed and severe violations may take place even
in large-scale simulations \cite{bcc-fcc-diamond-Kc,ferrenberg2018}.
Besides, FOPTs are arguably simpler than the continuous transitions
\cite{wilson,fisher_scaling_1982,maxwell_construction} so the task of
describing FOPTs within NPRG approach seems to be more feasible than
the incorporation of realistic critical behaviour into, e.g., the cluster
methods \cite{ngCVM,tokar_new_1997,tan_topologically_2011} efficient in
the description of FOPTs.

This viewpoint was implemented in \cite{maxwell_construction}
with some success.  The RG equation used by the authors allowed
them to calculate with a good accuracy the critical temperature
and to eliminate the van der Waals loops that plague mean field (MF)
theories of FOPTs. This ensued in the correct prediction of the infinite
susceptibility in the coexistence region but the discontinuity in the
inverse susceptibility at the region boundaries in three dimensional (3d)
case could not be reproduced \cite{maxwell_construction}.  Besides, the
calculated critical exponents did not agree well with known
values.  Nevertheless, the results show that NPRG is a viable approach
to the problem of simultaneous description of the continuous phase
transitions and the FOPTs while the cause of difficulties encountered
in \cite{maxwell_construction} should be sought in deficiencies of
NPRG implementation because the approach in itself is exact \cite{wilson}.

The aim of the present paper is to apply to the problem of unified
description of phase transitions of all kinds the RG equation derived in
\cite{1984,tokar2019,sc-lpa-rg}. As in \cite{maxwell_construction},
it was based on the local potential approximation (LPA)
\cite{wilson,riedel_tricritical_1972,nicoll_exact_1976,BAGNULS200191,%
caillol_non-perturbative_2012,local_potential} but was derived
within different---``layer-cake''---renormalization scheme. The
approach proved to be efficient in calculating critical temperatures
and other non-universal quantities in 3d spin-lattice models
\cite{tokar2019,sc-lpa-rg}; besides, the critical exponents better agreed
with the known values than in \cite{maxwell_construction}.

In the present paper it will be shown that an advantage of the RG
equation of \cite{1984,tokar2019,sc-lpa-rg} is that it can
be cast in the form of the generalized Burgers' equation \cite{gbe}
with time-dependent viscosity. The FOPTs in this formalism appear
as the shock wave solutions when the equation becomes inviscid. This
results in the infinite susceptibility in the coexistence region and
in the appearance of the discontinuities of the inverse susceptibility
at the region boundaries.  The quantitative accuracy of the theory
will be tested on the spin-1 BCM on the simple cubic (sc) lattice
\cite{blume_theory_1966,capel_possibility_1966}.  This choice was
motivated by the fact that the solution of BCM in the MF approximation
(see Appendix A in \cite{butera_blumecapel_2018}) qualitatively
reproduces the phenomenology of the extended Landau-Ginzburg model
\cite{landau_statistical_1980}. The latter along with the usual quartic
($\phi^4$) term \cite{landau_statistical_1980,wilson} contains the sixth
order term ($\phi^6$) which makes possible to describe in addition to
the conventional critical behaviour also the tricritical point and the
FOPTs. For our purposes the advantage of BCM in comparison with the
Landau-Ginzburg model is that because of the spin-lattice nature of BCM
there exists a wealth of quantitative data obtained in MC simulations
and in TSE (see extensive bibliography and TSE solutions in 
\cite{butera_blumecapel_2018}) needed in quantitative comparison.
\section{The model and the RG formalism}
In notation of  
\cite{capel_possibility_1966,deserno_tricriticality_1997,ozkan_critical_2006} 
the Hamiltonian of the ferromagnetic BCM in zero external field reads
\begin{equation}
	{\cal H} = -J\sum_{\langle ij\rangle}{ s_is_j}+D\sum_i{ s_i^2}
	\label{HI}
\end{equation}
where $J>0$ is the interaction between nearest neighbour (nn) spin-1
Ising spins; subscripts $i,j$ denote the sites of the sc lattice
of size $N$; the summation in the first term on the right hand side
(r.h.s.) is over nn pairs and $D$ is the crystal field splitting
\cite{blume_theory_1966,capel_possibility_1966} or the anisotropy field
\cite{butera_blumecapel_2018,ozkan_critical_2006}.

To simplify notation, (\ref{HI}) is convenient to cast in the form
\begin{eqnarray}
	&&H=\frac{\cal H}{T}=-K\sum_{<ij>}s_is_j+\Delta\sum_is_i^2\nonumber\\
	&&=\frac{1}{2}\sum_{ij}(\hat{\epsilon}_{ij}+r\delta_{ij})s_is_j
+\left(\Delta-\frac{qK+r}{2}\right)\sum_is_i^2.
	\label{H}
\end{eqnarray}
where temperature $T$ is assumed to be measured in energy units:
$T=k_BT^{(\rm K)}$ ($T^{(\rm K)}$ the temperature in Kelvins).
Besides, $J$ will be chosen as the energy unit so in (\ref{H})
$K=1/T$ and $\Delta=D/T$. Matrix $\hat{\epsilon}$ in (\ref{H})
is defined as
\begin{equation}
	\hat{\epsilon}=\big[qK\delta_{ij}-K_{ij}\big],
	\label{epsilon}
\end{equation}
where $q$ is the coordination number equal to 6 for the sc lattice,
$K_{ij}$ is equal to $K$ when $i$ and $j$ are nn sites and to zero
otherwise. The form of $\hat{\epsilon}$ in (\ref{epsilon}) has been
chosen for convenience of comparison with the RG theory dealing with
the Fourier components \cite{wilson} in the long wavelength limit
\begin{equation}
\epsilon({\bf k})_{k\to0}\propto k^2
	\label{k20}
\end{equation}
where ${\bf k}$ is the lattice Fourier momentum.  With similar purpose
an arbitrary parameter $r$ has been introduced in (\ref{H}) in order
that the inverse of the matrix in the parentheses in the first term on
the second line in (\ref{H}) $\hat{G}=(\hat{\epsilon}+r)^{-1}$ under
the lattice Fourier transform took the form \cite{tokar2019,sc-lpa-rg}
\begin{equation}
	G({\bf k})=\frac{1}{\epsilon({\bf k})+r}.
	\label{G(k)}
\end{equation}

The fluctuating field is introduced in the formalism by replacing the
trace over the discrete spins $s_i=0,\pm1$ in the BCM partition function
by $N$-dimensional integral over the continuous spins $-\infty<s_i<\infty$
as
\begin{equation}
	Z[h]=\int\cdots\int e^{-H +\sum_i h_is_i}
	\prod_lds_l[\delta(s_l)+ 2\delta(s_l^2-1)],
       \label{Z}
\end{equation}
where interaction with the source field $h$ has been separated from
Hamiltonian because it will be necessary for derivation of thermodynamic
relations, in particular, in the Legendre transforms.

Next with the use of $N$-dimensional Gaussian integral and the shift
operator \cite{gamma_exp,tokar_new_1997,sc-lpa-rg,tokar2019} the partition
function can be cast in the form
\begin{equation}
	Z[h]=\exp\left(\frac{1}{2}h\hat{G}h\right)R[\hat{G}h]
	\label{Z2}
\end{equation}
where
\begin{equation}
	R[s]=
	\exp\left(\frac{1}{2}{\partial_s}\hat{G}
{\partial_s}\right)\exp\left(-\sum_iu^{b}(s_i)\right).
	\label{R}
\end{equation}
The unrenormalized (``bare'') local potential in this expression is
defined as
\begin{eqnarray}
&&\exp\left[-u^b(s_i)\right]=[\det(2\pi\hat{G})]^{1/2N}\nonumber\\
&&\times\exp\left[-\left(\Delta -\frac{qK+r}{2}\right)s_i^2\right]
[\delta(s_i)+ 2\delta(s_i^2-1)]
	\label{ub}
\end{eqnarray}
where use has been made of (\ref{H}) and (\ref{Z}).
\subsection{Self-consistent RG equation} 
\begin{figure}[htp]
\centering \includegraphics{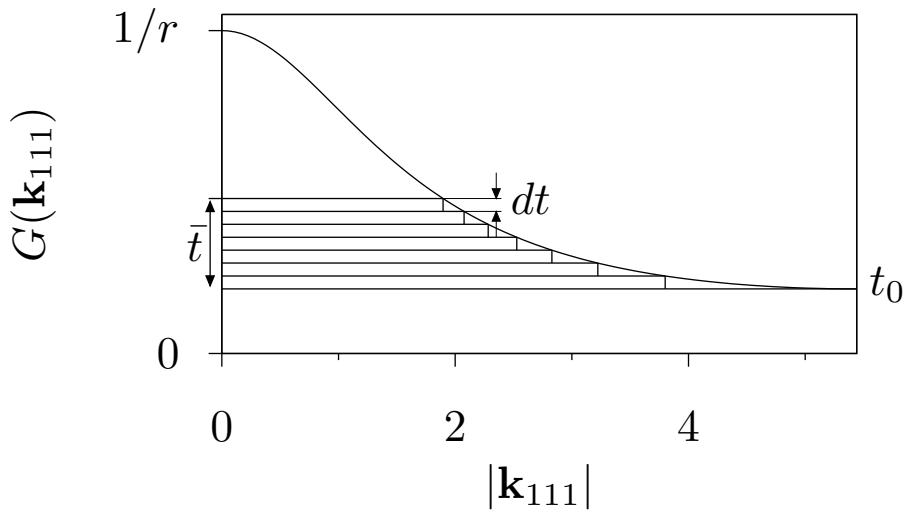}
\caption{\label{figure1} Illustration of the layer-cake renormalization
scheme. The layers are cross-sections of $dG=\theta[G({\bf k})-t]dt$
along the diagonal of the Brillouin zone on which $G({\bf k})$ reaches
its minimum value $t_0=1/(2qK+r)$; $\bar{t}= t-t_0$.}
\end{figure}

Because the bare potential in
 (\ref{ub}) is site-local, the BCM can be solved by the SC-LPA RG
approach \cite{sc-lpa-rg,tokar2019} based on the differential equation
\begin{equation}
        u_t = \frac{1}{2}\left[p(t)u_{xx}
        - u_x^2\right].
       \label{LPA}
\end{equation}
where $u(x,t)$ is the local potential that depends on the evolution
parameter $t$ which defines the stage of the ``layer-cake'' renormalization
(see figure \ref{figure1}) and on the local field $x$. The subscripts
$t$ and $x$ denote partial derivatives and
\begin{equation}
        p(t)=\rho_{tot}(t^{-1}-r)=\int_0^{t^{-1}-r}dE \rho(E),
       \label{p-latt}
\end{equation}
where $\rho(E)$ is the density of states of the quasiparticle band with
dispersion $\epsilon({\bf k})$. The initial value of $u(x,t=0)$ is given
by $u^b(x)$ in (\ref{ub}). It is singular because of the delta-functions
which is inconvenient for numerical solution of (\ref{LPA}). Fortunately, 
$p(t)=1$ in the range $0\leq t\leq t_0$, where
\begin{equation}
	t_0=\min G({\bf k})=1/(r+E_{max})=1/(r+2qK),
	\label{t0}
\end{equation}
(see figure \ref{figure1}) and $E_{max}=\max_{\bf k}\epsilon({\bf k})$
is the upper edge of spectrum $\rho(E)$.  It is easy to see that when
$p$ is constant, Eq. (\ref{LPA}) turns into a linear diffusion equation
for $\exp[-u(x,t)$. The equation is solved exactly with the use of the
Gaussian diffusion kernel \cite{1984,sc-lpa-rg,tokar2019} and at $t=t_0$
the solution reads
\begin{eqnarray}
	&&e^{-u(x,t_0)}=(2\pi t_0)^{-1/2}\int dx^\prime e^{-(x-x^\prime)^2/2t_0}e^{-u^b(x^\prime)}
	\nonumber\\
	&&=Ce^{-x^2/2t_0}\left(1+2e^{-(\Delta+qK/2)}\cosh\frac{x}{t_0}\right)
	\label{w}
\end{eqnarray}
where $C=\det(\hat{G}/t_0)^{1/2N}$.

The partition function (\ref{Z}) for homogeneous field $h$ or,
equivalently, the dimensionless FE per site is expressed in
terms of the local potential in the LPA as
\begin{equation}
	f(h) = -\ln Z(h)/N\simeq \left[u^R(x)-rx^2/2\right]_{x=h/r}.
	\label{f}
\end{equation}
Here and throughout the paper superscript ``$R$'' will mean that the quantity
is fully renormalized. In the case of the RG solution this will mean that its
``time'' coordinate $t=t^R=1/r$ is at the endpoint of the RG trajectory
(see figure \ref{figure1}).

The accuracy of (\ref{LPA}) can be improved by an appropriate choice of
the arbitrary parameter $r$. In \cite{sc-lpa-rg,tokar2019} it was shown
that good results can be obtained with the self-consistency (SC) condition
consisting in the choice of $r$ in such a way that it approximated the
exact mass operator of the model in the long-wavelength limit
\begin{equation}
	G^R({\bf k\to0})\approx G({\bf k\to0})=1/r.
	\label{G=G}
\end{equation}
In terms of the local potential the condition reads
\cite{sc-lpa-rg,tokar2019}
\begin{equation}
	\left.u^R_{xx}\right|_{h=0}=0.
	\label{SC}
\end{equation}
In this equation the external field is supposed to be equal to
zero because in the present paper mainly this case will be of
interest. However, in the BCM the phase transitions below $T_{tr}$
occur also at the ``wings'' \cite{blume_ising_1971} at $h\not=0$. In
this case condition (\ref{SC}) should be correspondingly modified.

As explained in \cite{sc-lpa-rg,tokar2019} and will be further
discussed in section \ref{fopts} below, in the coexistence region
 (\ref{LPA}) produces singular solutions difficult to deal with
numerically.  To overcome the problem, auxiliary local field variable $y$
and auxiliary local potential $v(y,t)$ were introduced via a Legendre-like
transform
\begin{eqnarray}
 \label{h}
  && x = y+\bar{t}  v_y \\ 
 \label{u(y)}
  && u = v+\bar{t}  v_y^2/2
\end{eqnarray}
where $\bar{t}=t-t_0$. In these variables (\ref{LPA}) takes the form
\begin{equation}
	v_t = \frac{p(t)v_{yy}}{2(1+\bar{t} v_{yy})}, \label{LPA2}
\end{equation}
and the evolution parameter varies from $t_0$ to $t^R=1/r$. The initial
condition at $t=t_0$ is obtained from (\ref{w}) 
\begin{eqnarray}
        v_0(y) &=& \frac{y^2}{2t_0}
        -\ln\left(1+2e^{-(\Delta+3K)}\cosh\frac{y}{t_0}\right)\nonumber\\
&&-\frac{1}{2}\left(\frac{1}{N}\sum_{\bf k}\ln G({\bf k})-\ln t_0\right)
       \label{v0}
\end{eqnarray}
with the use of the fact that at $t=t_0$ $\bar{t}=0$ so in (\ref{h})
and (\ref{u(y)}) $x=y$ and $u(x,t_0)=v(y,t_0)=v_0(y)$.  The SC condition
formally remains as in (\ref{SC}) \cite{sc-lpa-rg,tokar2019}
\begin{equation}
	\left.v^R_{yy}\right|_{h=0}=0.
	\label{SC2}
\end{equation}
The physical field and the local potential are obtained from the
auxiliary quantities (\ref{h}) and (\ref{u(y)}) in parametric form
as \cite{sc-lpa-rg}
\begin{eqnarray}
 \label{h1}
  && x=h/r = y+\bar{t}^R  v^R_y \\ 
 \label{u1(y)}
  && u^R = v^R+\bar{t}^R (v^R_y)^2/2
\end{eqnarray}
Equations (\ref{LPA2})---(\ref{SC2}) will be called the SC-LPA RG
equations.  They will be solved numerically throughout the paper by
numerical procedure described in \cite{sc-lpa-rg,tokar2019}.
\section{Continuous transitions in BCM}
As was pointed out in Introduction, continuous phase transitions
in BCM are located on three lines terminating at TCP.  However,
two of them bound the side ``wings'' appearing at $h\not=0$
(see figure 1 in \cite{butera_blumecapel_2018}) while in the
present paper we will restrict consideration mainly to case
$h=0$ which was thoroughly investigated within MC and TSE methods
\cite{deng_red-bond_2004,ozkan_critical_2006,hasenbusch_finite_2010,%
fytas_universality_2013,butera_blumecapel_2018} and our aim in the present
paper is to validate present theory via its quantitative comparison
with the most reliable alternative approaches.

\begin{figure}[htp]
\centering \includegraphics{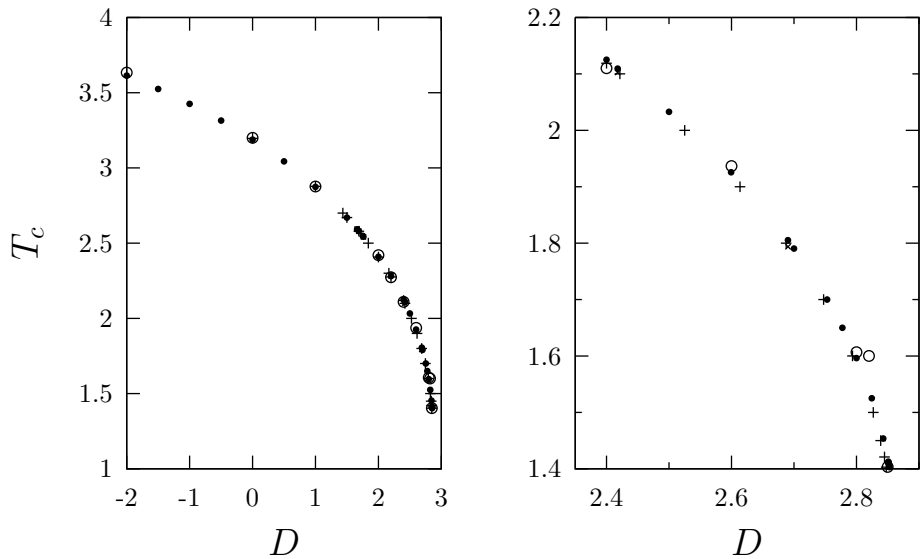}
	\caption{\label{figure2} Comparison of the critical
temperature of continuous phase transitions in the BCM calculated
for different values of the anisotropy $D$ by different techniques:
$\bullet$---present study; $\circ$---numerical simulations of
\cite{ozkan_critical_2006}; $\times$---Monte Carlo simulations of
\cite{hasenbusch_finite_2010}; $+$---high temperature expansions of
\cite{butera_blumecapel_2018}.	The lowermost points correspond to TCP
(somewhat different in different approaches).}
\end{figure}

At $h=0$ the critical line extends from $D=-\infty$ where BCM coincides
with the spin-1/2 Ising model to $D_{tr}$ corresponding to TCP, as shown
in figure \ref{figure2}. The critical temperature $T_c$ at each $D$
has been found as the temperature at which the SC value of $r$ turned
to zero (for details see \cite{sc-lpa-rg,tokar2019}). As can be seen,
$T_c$ predicted by the SC-LPA RG equation are in good agreement with the
results obtained in numerical simulations and in TSE.  Quantitatively the
agreement can be characterized by comparison with six precision values of
$T_c$ presented in Table VI of \cite{hasenbusch_finite_2010} and with
$T_c$ for $D=0$ from \cite{fytas_universality_2013}. The discrepancy
varied in the narrow range in 0.02\%--0.6\% and taking into account
that at $D=-2$ in figure \ref{figure2} $T_c\simeq3.61$, that is, in the
interval $-\infty<D<-2$ $T_c$ changes less than on 30\% and that at the
interval ends the SC-LPA predictions are off not more than on 0.6\%,
it seems plausible that this accuracy would hold for all values of $D$
for the continuous transitions at $h=0$.
\subsection{TCP}
As mentioned in the introduction and discussed in detail in
\cite{riedel_scaling_1972}, TCP is the point of intersection of three
critical lines and of one FOPT line. It means that, on the one hand,
the behaviour of thermodynamic quantities in the vicinity of TCP is
very complex but, on the other hand, there should be many ways of
determining its position on the phase diagram. For example, when
moving along a critical line the critical exponents will undergo
an abrupt change at TCP from their Ising values (in the BCM case)
to the TCP values. In particular, the exponent $\eta$ jumps from
$\eta\approx0.04$ \cite{Pelissetto2002CriticalPA} to $\eta_{tr}=0$
\cite{riedel_tricritical_1972}. With the exception of $\eta_{LPA}$ which
is also zero, other LPA exponents undergo the jump because they differ
from their TCP counterparts.  Thus, TCP can be found as the point at the
$h=0$ critical line where $\gamma$ jumps from $\gamma_{LPA}\simeq1.3$ to
$\gamma_{tr}=1$ and/or where $\beta$ change its value from $\simeq0.325$
to 0.25.

However, similar to other numerical studies
\cite{ozkan_critical_2006,butera_blumecapel_2018}, in the SC-LPA it was
difficult to obtained the truly abrupt change in the exponent values due
to finite accuracy of the calculations.  The cause of the problem was
that on approach to TCP the critical region of the Ising universality
class shrinks to zero, as can be shown analytically, e.g., in the mean
field approximation \cite{butera_blumecapel_2018} which is qualitatively
correct for 3d TCP.  So due to the limited accuracy of the computations,
the critical behaviour on approach to TCP was not possible to study
at distances smaller than $\tau\lesssim4\cdot10^{-3}$.  Within this
distance the critical exponents have been fitted to the numerical data
with the use of effective exponents \cite{sc-lpa-rg} which exhibited a
transient behaviour by changing from the LPA to TCP values. Fortunately,
the interval of such behaviour was sufficiently narrow, in particular, it
was much narrower than in TSE approach in \cite{butera_blumecapel_2018},
so the TCP coordinates in figure \ref{figure2} determined in this way
\begin{equation}
	D_{tr}=2.853 \mbox{\ and\ } T_{tr}=1.403
	\label{DtTt}
\end{equation}
compare well with known estimates. For example, they differ only
on less than 0.1\% from the values found in MC simulations in 
\cite{deng_red-bond_2004} which in \cite{butera_blumecapel_2018}
were reckoned to be the most accurate among available to date.

An important advantage of the RG approach is that in contrast to MC
and TSE techniques that rely solely on their numerical accuracy,
the tricritical behaviour of the LPA RG equation can be analyzed
analytically. Because the LPA equation (\ref{LPA}) has been derived from
the RG equation in \cite{1984}, the linearized form of the latter can
be used to determine the TCP critical exponents in the LPA. According
to \cite{riedel_tricritical_1972}, at the TCP fixed point the local
potential in 3d (the upper critical dimension) is equal to zero  so
in the RG equation (12) of \cite{1984}  linearized around the fixed
point we should set (in the notation of that reference) $d=3$ $z^*=0$
(hence, $r=0$), $\beta=(d-2)/8=1/8$ and $n=1$ (the Ising universality
class) to get the equation defining the eigenfunctions $e_j$ and the
eigenvalues $Y_j$ of the perturbations near the TCP fixed point:
\begin{equation}
	-e_j^{\prime\prime}+\bigg(\frac{x^2}{16}-3\frac{1}{4}\bigg)e_j=-Y_je_j.
	\label{eq12}
\end{equation}
With the use of operators $a^\pm = 8^{-1/2}x\pm2^{1/2}d/dx$ the equation can
be cast in the form
\begin{equation}
	\frac{1}{2}\hat{n}e_j=(3-Y_j)e_j,
	\label{a+a-}
\end{equation}
where $\hat{n}=a^+a^-$ is the occupation number operator with the eigenvalues
0,1,2,\ldots which can be used as the index $j$. Thus,
\begin{equation}
	Y_j = 3-j/2
	\label{Yj}
\end{equation}
in accordance with known exact values
\cite{riedel_tricritical_1972,stinchcombe,DtTt2004}.  As is seen, the
large eigenvalues that define the tricritical behaviour are those
with smaller $j$. The first three of them are $Y_0=3$, $Y_1=5/2$ and
$Y_2=2$. The largest eigenvalue $Y_0$ simply reflects
the fact that in 3d the FE of the system grows as the third power
of the linear size. The non-trivial eigenvalues define the critical
exponent $\nu_{tr}=1/Y_2=0.5$ and $\beta_{tr}=(3-Y_1)/Y_2=0.25$
\cite{wegner_corrections_1972,deng_simultaneous_2003}. Now
using the scaling relation $\gamma = \nu(2-\eta)$ (valid also
at TCP \cite{riedel_tricritical_1972}) with $\eta_{tr}=0$ one gets
$\gamma_{tr}=1$ in compete agreement with the known exact values. 

Determining TCP coordinates (\ref{DtTt}) through the jump in the
critical exponents is, arguably, the most difficult way to proceed. An
easier way would be to determine them by finding such $D_{tr}$ that for
$D>D_{tr}$ the critical points with $h,r=0$ disappear.  Several more
methods exist in the FOPTs region. (i) TCP can be found as the point
defining the start of the FOPT transition line, that is, the point
at which the quartic term in renormalized local potential changes its
sign \cite{landau_statistical_1980}, that is, when $u^R_{xxxx}=0$. (ii)
Further, the difference of FOPTs from the continuous transitions is that
the spontaneous magnetization $m_0$ (the subscript is for $h=0$ condition)
at $T_c$ (we will keep this notation also for FOPT temperatures) is
finite.  So TCP can be found as the point where $m_0$ turns to zero on
approach to TCP from below $T_{tr}$. (iii) Yet another way of determining
TCP can be based on the fact that the latent heat of FOPTs disappears
at TCP. All these possibilities have been tested in calculations below
and deviations from (\ref{DtTt}) have been found to be $\Or(0.1\%)$.
\section{\label{fopts}FOPTs as the shock wave solutions of the RG equation}
In continuous transitions the critical temperature can be found as the
point of divergence of the correlation length which in our formalism
means $r=0$.  Alternatively, it can be determined as the point where
the order parameter $m_0$ switches from zero to non-zero values.

At the FOPTs, however, both $r$ and $m_0$ remain finite so usually
$T_c$ is defined as the temperature at which the FE curves $f_o(T)$ and
$f_d(T)$ calculated, respectively, in the ordered and in the disordered
phases intersect.  This implicitly presumes that $f_o$ and $f_d$ can
be continued inside another phase. However, as pointed out in 
\cite{butera_blumecapel_2018}, this approach is controversial because
thermodynamic quantities at the FOPT point are singular, though
a detailed structure of the singularities is unknown (see 
\cite{zia_variety_1981,privman1982} and references therein).

\begin{figure}[htp]
\centering \includegraphics{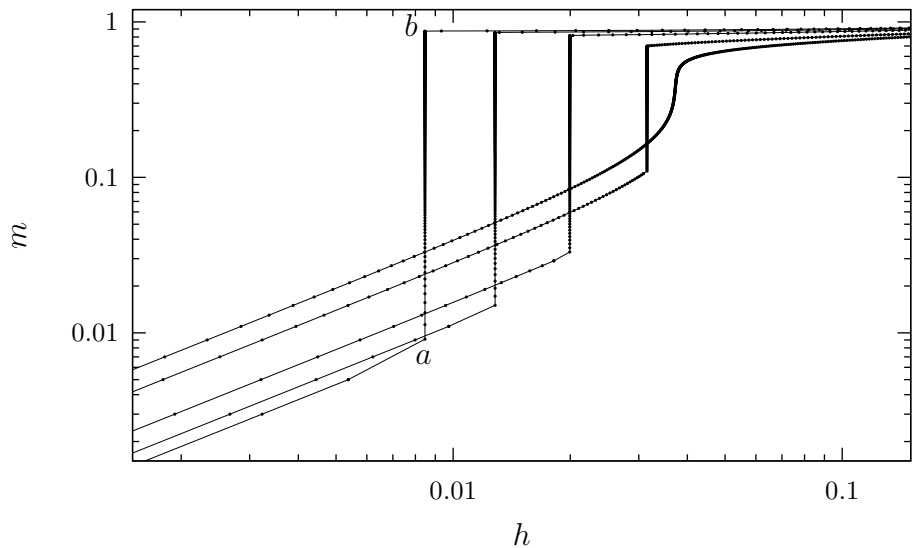}
\caption{\label{figure3} The equation of state ($m(h)$ isotherms) in the
wing region for $D=2.95$ and temperatures (from right to left) $T$=
1.27, 1.2, 1.1, 1.05 and 1.022 The calculated points are connected 
by lines for clarity. The rightmost curve corresponds to
the temperature just above the wing critical temperature $T_c^{\rm(wing)}$.}
\end{figure}

But in the BCM the FOPT line at $h=0$ is the intersection of two symmetric
surfaces of FOPTs called wings on which $h\not=0$ which join at $h=0$
(see, e.g., figure 1 in \cite{butera_blumecapel_2018}).  This can
be used to determine the line of FOPTs at $h=0$ as follows. Let us
for definiteness consider the wing at positive $h$. At a fixed value of
$D_{tr}<D<3$ and $T$ such that $T_c(h=0)<T<T_{tr}$ ($T_c(h=0)$ is the FOPT
temperature of interest) the line parametrized by $h>0$ at some point
$h_c$ will pierce the wing. In crossing the surface the order parameter
will abruptly change its value from, say, $m_a$ to $m_b$ (or from
$-m_a$ to $-m_b$ for negative $h$). At the $m(h)$ isotherm this will be
seen as the vertical line connecting two endpoints with different $m$
but the same $h$, as can be seen in figure \ref{figure3}. The equation of
state describing the lines in the figure has been obtained in parametric
form by using equation (35) from \cite{sc-lpa-rg} 
\begin{equation}
	m = -df/dh = y-t_0v^R_y
	\label{m0}
\end{equation}
for magnetization and (\ref{h11}) for $h$.  At $T_c(h=0)$ both $h_c$
and $m_a$ turn to zero, so any of these quantities may be used as the
order parameter of the transition.
Similar to the $h=0$ spin-1/2 Ising model studied in  
\cite{sc-lpa-rg,tokar2019}, in 
BCM in the coexistence region at $h=h_c\not=0$ the auxiliary potential
coincides with the particular Gaussian solution of (\ref{LPA2}) 
of the form
\begin{equation}
	v_G(y,t) = -\frac{(y-h_c/r)^2}{2\bar{t}^R }+\mbox{(f.i.t.)},
	\label{v_G}
\end{equation}
where (f.i.t.) denotes field-independent terms. Substitution of $v_G$
into (\ref{h}) shows that solution (\ref{v_G}) corresponds to a
constant external field $h=h_c$, that is, to the vertical lines in figure
\ref{figure3}.  Magnetization calculated from $v_G$ with the use of 
(\ref{m0}) coincided within the accuracy of the calculations with $m$
found from the numerical solution $v^R(y)$ in the interval $(y_a,y_b)$
corresponding to the interval $(m_a,m_b)$. This is analogous to the case
of the spin-1/2 Ising model \cite{sc-lpa-rg} where the Gaussian solution
described the behaviour below $T_c$ on the interval $(-m_0,m_0)$. This
seemingly universal behaviour was first observed in solutions of an LPA
RG equation in \cite{maxwell_construction} and was tentatively
attributed to a fixed point describing FOPTs. To clarify the issue in
the present context it is instructive to switch from the auxiliary to
the original variables $x$ and $u$.
\subsection{LPA RG equation as generalized Burgers' equation}
With the use of (\ref{f}) magnetization can be found as 
\begin{equation}
	m(h) = \frac{h}{r}-\left.\frac{u_x^R}{r}\right|_{x=h/r}.
	\label{m(h)}
\end{equation}
The first term on the r.h.s.\ describes the linear response 
to the external field while the second term may be called the non-linear
response because due to the SC condition (\ref{SC}) this term 
contributes to $m$ only in higher orders in $h$. Hence, the discontinuities
in $m(h)$ dependence may come only from this term. By differentiating  
(\ref{LPA}) with respect to $x$ it is easy to see that
\begin{equation}
	\mu(x)=u_x(x)
	\label{mu}
\end{equation}
satisfies the generalized Burgers' equation \cite{gbe}
\begin{equation}
	\mu_t+\mu\mu_x=(p/2)\mu_{xx}	
	\label{burgers}
\end{equation}
which coincides with the conventional viscous Burgers' equation (BE)
\cite{whitham_linear_2011} when the ``viscosity'' $p/2$ is constant.  But in
our case $p$ is a function of ``time'' $t$ and in \cite{gbe} $p(t)$
was interpreted as the cross section of a duct. Despite being dependent
on $t$, $p(t)$ is positive so can play the role of the diffusion term and
reproduce qualitatively the behaviour of the BE solutions. Furthermore,
because $p(t=t^R)=0$, it may be expected that fully renormalized $\mu^R$
would behave similar to the inviscid BE in the zero-viscosity limit,
in particular, exhibit the discontinuities seen in figure \ref{figure3}.
This possibility will be discussed in more detail in section \ref{ffp}.

The RG equation in auxiliary variables for $\mu(y,t)=v_y(y,t)=\mu(x,t)$,
where the last equality follows from equation (B.3) in \cite{sc-lpa-rg}
\begin{equation}
	u_x=v_y,
	\label{uxvy}
\end{equation}
is straightforwardly derived from (\ref{LPA2}) as
\begin{equation}
	\mu_t = \frac{p(t)\mu_{yy}}{2(1+\bar{t} \mu_{y})^2},  
	\label{LPA21}
\end{equation}

In principle, (\ref{LPA}) could be replaced by (\ref{burgers}) without
loss of generality because local potential $u$ needed in the expression
for FE (\ref{f}) could be obtained by simple integration of $\mu$ from,
say, 0 to $x$. The integration constant could be recovered from (\ref{w})
augmented by integration over $t$ of (\ref{LPA}) at $x=0$ as
\begin{equation}
\Delta u^R(0)=\frac{1}{2}\int_{t_0}^{t^R}dt[p(t)\mu_x(0,t)-\mu^2(0,t)].
	\label{delta_u}
\end{equation}
However, (\ref{LPA21}) implicitly contains the third derivative
$v_{yyy}$ which in view of the discontinuous second derivative would
introduce delta-function singularities into the equation causing
difficulties in numerical solution. Besides, the derivation of GBE
(\ref{burgers}) by simply taking the derivative over $x$ of (\ref{LPA})
would not work for Landau-Ginzburg models with multicomponent
vectors ($n>1$). The SC-LPA RG equations for such models derived in
\cite{tokar2019} directly generalize (\ref{LPA}) to $n$-component
case. When $p=0$ they acquire the form of the Hamilton-Jacobi
equations and in this form also can be used for the description of
FOPTs \cite{choquard_mean_2004,lorenzoni_exact_2019}. Therefore,
all numerical calculations in this paper have been done with the
use of equations for $u$ and $v$.  BE in different forms have been
used only to clarify the shock wave features of the solutions
because it is a conventional framework for such analysis
\cite{whitham_linear_2011,shock-waves,Brankov_1983,choquard_mean_2004}.
\subsection{Infinite range Ising model}
To gain qualitative insight into (\ref{burgers}), let us apply
it to a simple case where most calculations can be performed
analytically.  The infinite-range Ising model (IRIM) also known as
the Husimi-Temperley \cite{Brankov_1983} and the Curie-Weiss model
\cite{salinas_introduction_2001} can be solved exactly within the MF
theory for all values of the model parameters with the exception of the
coexistence region where the unphysical van der Waals loops appear. In
\cite{Brankov_1983,choquard_mean_2004} it was shown how they can be
avoided with the use of BE in the solution of the model. Incidentally,
IRIM can be also solved exactly within the SC-LPA RG approach. Therefore,
below we will show how equation (\ref{burgers}) can be used in the
solution of the problem of the van der Waals loops within the RG approach.

Similar to the Landau MF theory \cite{landau_statistical_1980}, IRIM
is a structureless model because when a spin interacts with the same
strength with any other spin in the system it does not matter how far
apart the spins are, what is the lattice they are on or whether it at
all exists. Therefore, in matrix $\hat{\epsilon}$ in (\ref{epsilon})
we do not specify the topology but formally set the coordination number
$q=N-1$ and define interactions between spins at different sites to be
$J/N$ where $J=\Or(1)$. From the requirement $\sum_j\hat{\epsilon}_{ij}=
0$ equivalent to (\ref{k20}) on any lattice it is easy to find
\begin{equation}
	\hat{\epsilon}^\infty = K(\hat{I}-\hat{E})
	\label{eps-infty}
\end{equation}
where $K=J/T$ is formally the same as in (\ref{epsilon}), $\hat{I}$
is the unit matrix and in matrix $\hat{E}$ all matrix elements are
equal to $1/N$. Matrix $\hat{E}$ is idempotent ($\hat{E}^2=\hat{E}$)
and so such is $\hat{I}-\hat{E}$.  The spectrum of idempotent matrices
consists of only two points: 0 and 1, so the two values of the spectrum
of $\hat{\epsilon}^\infty$ are 0 and $K$. The corresponding density of
states can be found as $\rho(E)=\Im G_{ii}(E+\mathrm{i}\varepsilon)/\pi$.
So using easily derivable expression
\begin{equation}
	-\left(\frac{1}{z-\hat{\epsilon}^\infty}\right)_{ii}
	=-\frac{1}{N}\frac{1}{z}-\left(1-\frac{1}{N},
	\right)\frac{1}{z-K}
	\label{Gii(z)}
\end{equation}
with $z=E+\mathrm{i}\varepsilon$ (the minus signs before all terms have been
introduced for compatibility with (\ref{G(k)})), one gets
\begin{equation}
	\rho^\infty(E) = N^{-1}\delta(E)+\delta(E-K)
	\label{rho}
\end{equation}
where in the second term $1/N\ll1$ have been neglected.  The upper edge
of the spectrum $E_{max}=K$ fixes the lowest value of the evolution
parameter $t$ defined in (\ref{t0}) as
\begin{equation}
	t^\infty_0=1/(r+K) 
	\label{t0-infty}
\end{equation}
The upper integration limit corresponds to $\max G=1/(r+E_{min})$ (see
figure \ref{figure1}) with $E_{min}$ always equal to zero in the present
formalism.  Thus, the total density of states needed in (\ref{p-latt}) is
\begin{equation}
	\rho^\infty_{tot}(E)=N^{-1}+\theta(E-K).
	\label{rho-infty}
\end{equation}
Using (\ref{p-latt}) and (\ref{rho-infty}) it is easy to see
that in the whole integration range $t^R<t<t^\infty_0$ 
\begin{equation}
	p^\infty(t)=\rho^\infty_{tot}(t^{-1}-r)=N^{-1}.
	\label{p-infty}
\end{equation}
Thus, GBE (\ref{burgers}) in this case takes the form of the conventional
BE with constant viscosity $1/2N$
\begin{equation}
	\mu^\infty_t+\mu^\infty\mu^\infty_x=(1/2N)\mu^\infty_{xx}	
	\label{burgers-infty}
\end{equation}
and coincides with equation (3.2) derived in \cite{Brankov_1983} in a
different approach. Now following that paper we could linearize BE with
the use of the Cole-Hopf substitution and analyze the solution along
the lines of \cite{Brankov_1983}. However, our aim is not to solve
IRIM by the RG technique but to show that the method of solution of
GBE (\ref{burgers}) used in this paper and \cite{tokar2019,sc-lpa-rg}
leads to the shock wave solutions at FOPTs similar to those suggested
in \cite{Brankov_1983,choquard_mean_2004} for IRIM.

To begin with, let us discuss, how the mechanism by which even an
infinitesimal viscosity in BE suppresses the multivalued solutions of
inviscid BE, can be understood in the framework of the SC-LPA RG approach.
To this end, let us first discard the viscosity in (\ref{burgers-infty})
by taking the naive thermodynamic limit $N\to\infty$ to obtain the
inviscid BE.  Now the corresponding equation (\ref{LPA21}) in the
auxiliary variables in the IRIM case reads
\begin{equation}
	\mu^\infty_t(y,t)=0.
	\label{mu_t}
\end{equation}
With the parameters corresponding to $\hat{\epsilon}^\infty$ the initial
condition for the spin-1/2 Ising model given in equation (A.14) of  
\cite{sc-lpa-rg} reads
\begin{equation}
	u^\infty_0(x)=\frac{x^2}{2t^\infty_0}
	-\ln\left(2\cosh\frac{x}{t^\infty_0}\right).
	\label{u-infty}
\end{equation}
Differentiation with respect to $x$ gives the initial condition at
$t=t^\infty_0$ for (\ref{burgers-infty}) as
\begin{equation}
\mu^\infty_0(x) = [x-\tanh(x/t^\infty_0)]/t^\infty_0.
	\label{mu0-infty}
\end{equation}
The initial $\mu^\infty_0(y)$ can be obtained from (\ref{mu0-infty})
by simply replacing $x$ in (\ref{mu0-infty}) with $y$ because as already
mentioned, according to (\ref{h}) and (\ref{u(y)}) the variables coincide
at $t=t_0$.  Further, because according to (\ref{mu_t}) $\mu^\infty(y)$
does not depend on $t$, we immediately obtain the solution in auxiliary
variables
\begin{equation}
	\mu^\infty(y)=[y-\tanh(y/t^\infty_0)]/t^\infty_0
	\label{solution}
\end{equation}
and similarly for the local potential 
\begin{equation}
	v^\infty(y)=\frac{y^2}{2t^\infty_0}
	-\ln\left(2\cosh\frac{y}{t^\infty_0}\right)
	\label{v-infty}
\end{equation}
(cf.\  with (\ref{u-infty})). The subscripts $0$ have been omitted in
the last two expressions because the solutions are valid at all $t$,
in particular, at $t=t^R$ which will be implicitly assumed everywhere
below unless stated otherwise.

\begin{figure}[htp]
\centering \includegraphics{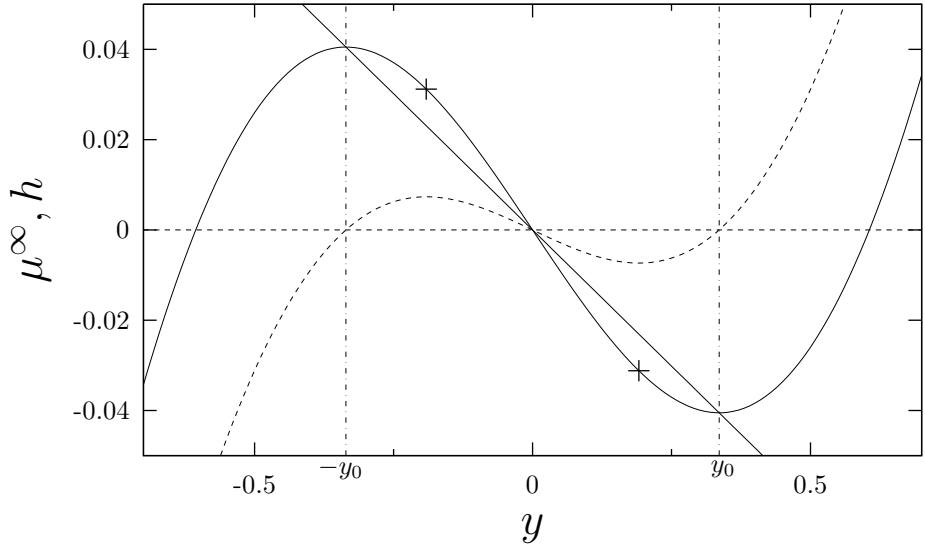}
\caption{\label{figure4}Plots of $\mu^\infty(y)$ (\ref{solution})
(solid curve) and $h(y)$ from (\ref{x-infty}) (dashed curve) that define
$\mu^\infty(h)$ for IRIM in parametric form at $K=1.05$. Straight solid
line is $\mu_G$ (\ref{mu_G}); dashed-dotted lines bound the coexistence
region, $y_0$ is from (\ref{m0-infty}); symbols (+) bound the segment
of $\mu^\infty$ curve where it is steeper than $\mu_G$.}
\end{figure}

As noted previously, due to (\ref{uxvy}) $\mu(x,t)=\mu(y,t)$ so to
determine $\mu^\infty(x,t^R)$ (\ref{solution}) should be augmented by
(\ref{h1}) which defines the dependence of $x$ on $y$ 
\begin{equation}
  x=h/r = y+\bar{t}^R \mu^R.
	\label{h11}
\end{equation}
With the use of (\ref{t0-infty}) and (\ref{solution}) (\ref{h11}) can
be conveniently written as
\begin{equation}
	h=y/t^\infty_0 -K\tanh(y/t^\infty_0).
	\label{x-infty}
\end{equation}.
Another measurable quantity---magnetization---is given by (\ref{m0})
which in terms of $\mu$ reads
\begin{equation}
	m = y - t_0\mu^R.
	\label{m}
\end{equation}
Substitution here of (\ref{mu0-infty}) gives
\begin{equation}
	m=\tanh(y/t^\infty_0).
	\label{m-infty-y}
\end{equation}
In conjunction with (\ref{x-infty}) it leads to the exact IRIM
MF equation \cite{salinas_introduction_2001}
\begin{equation}
	m=\tanh(Km+h)
	\label{MF}
\end{equation}
for determining the equilibrium magnetization in terms of temperature
and the external field. In particular, from (\ref{MF}) it is easy to
find that when $h=0$ the critical temperature $T_c=1$, as follows from
$1/T_c=K_c=1$.  Besides, by comparing the last two equations it is easy
to see that the spontaneous magnetization $m_0$ and corresponding to it
$y_0$ in figure \ref{figure4} are proportional to each other
\begin{equation}
m_0=y_0/Kt_0.	
	\label{m0-infty}
\end{equation}
Finally, expressing $u^R$ in (\ref{u1(y)}) through (\ref{v-infty}) and
$\mu^R=v^R_y$ and substituting this into FE (\ref{f}), where the last
term with the use of (\ref{h11}) should be expressed through $y$ and
$\mu^R$, one obtains an expression for FE in terms of $y$ and $\mu^R$.
Now noting that the argument of the hyperbolic cosine is the same as
in (\ref{MF}) and rearranging remaining terms so as to express them
through $m$ from (\ref{m}) one arrives at the exact expression for IRIM
FE \cite{salinas_introduction_2001,Brankov_1983}
\begin{equation}
	f^\infty = Km^2/2-\ln\left[2\cosh(Km+h)\right].
	\label{f-infty}
\end{equation}

The obtained solution of IRIM is exact everywhere except in the
coexistence region where it exhibits the unphysical van der Waals
loops characteristic of MF theories. In terms of $\mu^\infty(x,t)$
this can be seen from the graphs of $h(y)$ and $\mu(y)$ shown in
figure \ref{figure4}. For any value of $h$ in the range between the
two local extrema there are three values of $y$, hence, three values
of $\mu^\infty(y)$. So $\mu^\infty(h)$ is a three-valued function in
this range.

To see how the viscosity, even infinitesimal, can resolve the problem
of the van der Waals loops let us consider equation (\ref{LPA21}) as
a diffusion equation with the diffusivity depending on spatiotemporal
coordinates:
\begin{equation}
	{\cal D}(y,t)=\frac{p(t)}{2(1+\bar{t} \mu_{y})^2}
	\label{Ddiff}
\end{equation}
In the discussion below we will often omit references to IRIM because
the majority of arguments are expected to be valid in general case. The
qualitative picture presented below have been observed in the numerical
solutions of the SC-LPA RG equations for BCM and the Ising model.

Because we will be interested in the behaviour of the diffusivity in the
coexistence region where derivative $\mu_y$ is negative, to simplify the
discussion let us introduce the minus sign explicitly into the expression
in the parentheses in the denominator of (\ref{Ddiff}) as
\begin{equation}
	\Psi(y,\tau) = 1-\tau|\tilde{\mu}_y|
	\label{psi}
\end{equation}
where additionally the normalized quantities $\tau=\bar{t}/\bar{t}^R$ and
$\tilde{\mu}=\bar{t}^R\mu$ have been introduced.  Their convenience is
that during the evolution $\tau$ varies from 0 to 1 so from (\ref{psi})
it is obvious that $|\tilde{\mu}_y^R|$ cannot exceed unity because in
this case $\Psi^R$ would be negative which means that at some $\tau_c>0$
it crossed zero because $\Psi(y,0)=1>0$. But if $p\not=0$ this would
introduce in ${\cal D}$ nonintegrable singularity in $t$ which means
that the renormalization would not be possible to complete.  In fact, on
approach to $\tau_c$ the diffusivity will acquire arbitrarily large values
even for smallest viscosity. The enhanced diffusion will be smearing
$\mu(y,t)$ until making the slope $|\tilde{\mu}_y|$ smaller than unity.

Thus, the steepest $y$-dependence of $\mu^R$ compatible with viscosity is
$-1/\bar{t}^R$ which coincides with the slope of $\mu^R$ corresponding to
the Gaussian solution (\ref{v_G}). It has always been obtained in the
coexistence region in BCM and Ising models in the numerical solutions of
(\ref{LPA2}). The linear function
\begin{equation}
	\mu_G(y)=-\frac{y-h_c/r}{\bar{t}^R}
	\label{mu_G}
\end{equation}
for the symmetric case $h_c=0$ is drawn in figure
\ref{figure4}. Comparison with $t$-independent solution of IRIM
(\ref{solution}) shows that the latter cannot be valid at all $t$
because in the segment between the crosses $\mu^\infty$ is steeper than
$\mu_G$. Thus, with the inclusion of viscosity at least this segment
will be flattened by diffusion in the course of evolution, thus making
$\mu^\infty$ dependent on $t$.  In the case of IRIM it is easy to show
that at the end of evolution the initial $\mu^\infty(y)$ curve within
the segment in figure \ref{figure4} will coincide with $\mu_G$. The
evolution toward $\mu_G$ of the parts of the curve outside the segment
is not straightforwardly seen from the form of diffusivity (\ref{Ddiff}).
So instead the numerical procedure that has been used in the solution
of BCM has been applied to (\ref{LPA2}) for IRIM with $N=10^3$.  The
Gaussian potential (\ref{v_G}) corresponding to the thermodynamic limit
$N\to\infty$ has been obtained for $h_c=0$ in the whole coexistence
region with the accuracy better than 0.01\%.  When substituted in
(\ref{h11}) it gives $h=0$ within the region which means that
when $h$ crosses zero, the magnetization jumps from, say, $m_0$ to $-m_0$,
as expected. Thus, the RG flows during FOPTs in BCM and in the Ising
model are qualitatively the same as in the exactly solvable and well
understood IRIM \cite{Brankov_1983,choquard_mean_2004}.
\subsection{\label{ffp}FOPT fixed point}
The qualitative behaviour just described suggests that the
RG trajectory inside the coexistence region terminates with a
universal function (\ref{v_G}) which potentially is a candidate
for the discontinuity fixed point advocated in some approaches
\cite{nienhuis_first-order_1975,zia_variety_1981,fisher_scaling_1982,%
privman1982,maxwell_construction}. The attraction basin of the point
consists of Hamiltonians with initially renormalized $\mu_0(y)$ having
regions of $y$-dependence with negative slope exceeding that of $\mu_G$
or, equivalently, with the negative curvature of $v_0(y)$ larger in
absolute value than the curvature of $v_G$. The universal behaviour,
however, does not comprise the whole range of variation of the local
field $y$, as in continuous transitions. In BCM in the fully renormalized
case it is restricted to finite intervals in the auxiliary variables
$(y_a,y_b)$, where $y_{a(b)}$ correspond to magnetizations $m_{a(b)}$, in
the wing region at $h_c\not=0$ (see figure \ref{figure3}) or to $\pm m_0$
in the zero field case.  This apparently is a consequence of the finite
amplitude of fluctuations of the order parameter in FOPTs. In physical
variables in the fully renormalized case this range even shrinks to one
discontinuity point: $x_c=h_c/r$.

Discontinuities are crucial for the description of FOPTs so to get
physical insight into the RG flow leading to them in physical terms
let us approximate $\mu(x,t)$ for $t\to t^R$ as follows.  First,
let us assume that in the coexistence region and for $t$ close to
$t^R$ $\mu(y,t)\approx\mu_G(y)$.  Using $\mu(y,t)=v_y=u_x=\mu(x,t)$
and replacing in (\ref{h}) $v_y$ by $\mu(x,t)$ one finds that
$y=x-\bar{t}\mu(x,t)$.  Substituting this into the r.h.s.\ of (\ref{mu_G})
and replacing on the l.h.s.\ $\mu_G(y)$ by $\mu(x,t)$ one finds
\begin{equation}
	\mu(x,t)\approx-\frac{x-x_c}{t^R-t}\equiv \frac{\Delta x}{\Delta t}.
	\label{mu_Gx}
\end{equation}
As is seen, in the coordinates relative to the ``critical point''
$(x_c,t^R)$ the function is universal being independent of all parameters
of the system under consideration. Also, when $\Delta t\to 0$ the slope of
$x$-dependence tends to infinity, as is necessary for the susceptibility
in the coexistence region to be infinite. The range of validity of
(\ref{mu_Gx}), i.e., the coexistence region in physical coordinates, can
be easily assessed in the symmetric case $x_c=0$ using the fact that as
$t\to t^R$ and $|h|$ is small, according to (\ref{m(h)}) $\mu(x,t)\approx
-rm_0\mbox{sgn}(x)$ ($m_0>0$). But because $\mu(x,t)$ is a derivative
of a smooth function $u(x,t)$ it should be continuous so for small $x$
and $t\to t^R$ it can be modelled by a smeared sign function as
\begin{equation}
	\mu(x,t)\approx \left\{
		\begin{array}{ll}
			-rm_0\,\mbox{sgn}(x) & |x|\geq x_0\\
			\mu_G(x,t)|_{x_c=0} & |x|< x_0 \end{array}\right.
	\label{model-mu}
\end{equation}
where $x_0=rm_0(t^R-t)$. As can be seen, the universal fixed point in the
fully renormalized case $t=t^R$ reduces to the jump in $\mu^R(x)\simeq
-rm_0\mbox{sgn}(x)$ at a single point which simply corresponds to
switch of magnetization from $-m_0$ to $m_0$ when the external field
$h$ crosses zero (or from $m_a$ to $m_b$ when $h$ crosses $h_c$, as
in figure \ref{figure3}). The singular part of FE acquires the form
$f_{sing}\simeq m_0|h|$ which is easily guessed on phenomenological
grounds \cite{nienhuis_first-order_1975,fisher_scaling_1982}.  From the
RG standpoint, however, the result is not trivial because the singularity
in $u^R$ has been obtained by renormalization of the Landau-Ginzburg
Hamiltonian which by definition is analytic in the fluctuating field
(see (\ref{w})).

Thus, to adequately describe FOPTs it seems sufficient to
have $\mu^R(y,t)=\mu_G(y)$, or, equivalently, $v=v_G$ inside the
coexistence region.  However, there is a caveat.  From mathematical
literature \cite{whitham_linear_2011,shock-waves} it is known that
infinitely steep shock waves may occur only in the inviscid BE and
thus are incompatible with viscosity. In physical terms this can be
seen by substituting (\ref{mu_Gx}) in (\ref{delta_u}). Because of
the nonitegrable singularity, the contribution to $u$, hence, to FE
through (\ref{f}), diverges while FE should be finite and $\Or(1)$. In
the IRIM case $p=1/2N$ so in order the integral of the first term in
(\ref{delta_u}) was finite and $\Or(1)$ the derivative $\mu^R_x(0,t^R)$
should be $\Or(N)$, as indeed was found in \cite{Brankov_1983}. But
this is incompatible with (\ref{mu_Gx}).

In IRIM the necessary discontinuity is recovered in the thermodynamic
limit $N\to\infty$ but in short-range models $p(t)$ is nonzero for all $t$
except at $t=t^R$ where the integration range in (\ref{p-latt}) shrinks
to zero. This would be enough to make the integral in (\ref{delta_u})
convergent but may be insufficient for the appearance of the sharp shock
waves in GBE (\ref{burgers}). In such waves we expect $\mu(x,t)$ to
acquire the form of $\mu(x,t)$ in (\ref{mu_Gx}).  But when substituted
in (\ref{burgers}) the denominator of $\mu(x,t)$ will effectively
differentiate $p(t)$ at $t^R$ so the r.h.s.\ of the RG equations
will become effectively proportional to $p^\prime(t^R)$.  The latter
is proportional to $\rho(E=0)$ which is zero in the nn models that we
have been dealing with, so (\ref{burgers}) would approach the inviscid
BE for such models at the end of the evolution. This observation have
been checked by application of the SC-LPA RG equation to the solution of
a test model with $\rho(E)\not=0$ and no transitions have been found.
This fact may be the cause of the poor performance of LPA in 2d case.
The problem may disappear if momentum-dependent corrections to $r$
would make $\rho(E=0)$ equal to zero. This, however, will require going
beyond LPA.
\section{FOPTs in zero external field} 
In previous section the regions of the phase diagram in the wings
that appear in BCM below $T_{tr}$ have been briefly discussed. Their
detailed description in the MF approximation can be found in many papers
\cite{blume_theory_1966,blume_ising_1971,capel_possibility_1966,%
butera_blumecapel_2018} but reliable MC or TSE data seems
to be absent, so for the quantitative comparison we restrict
our consideration to much better studied zero-field case
\cite{deserno_tricriticality_1997,deng_red-bond_2004,ozkan_critical_2006,%
hasenbusch_finite_2010,butera_blumecapel_2018}.

Besides the vanishing of $h_c$ and/or $m_a$ suggested in previous section
as the ways of determining $T_c$ of the zero-field FOPTs in BCM, other
methods can be used. As noted earlier, conventionally FOPTs are defined as
the points where the values of free energies in two phases coincide, i.e.,
where $f_d=f_o$. The peculiarity of the numerical SC-LPA RG solutions
was that they were unique in the sense that it has been impossible to
continue the disordered solution below $T_c$ and the ordered one above
$T_c$. This suggests yet another way of determining the transition
temperature as the point where the ordered solution disappears and the
single coexistence region below $T_c$ splits into two wing coexistence
regions, as in figure \ref{figure3}. This method has appeared to be the
most practical because the temperature step could be chosen sufficiently
small to guarantee desired accuracy. $T_c$ value has been placed at the
middle point between the ordered solution at lower temperature and the
disordered solution one step above.

\begin{figure}[htp]
\centering \includegraphics{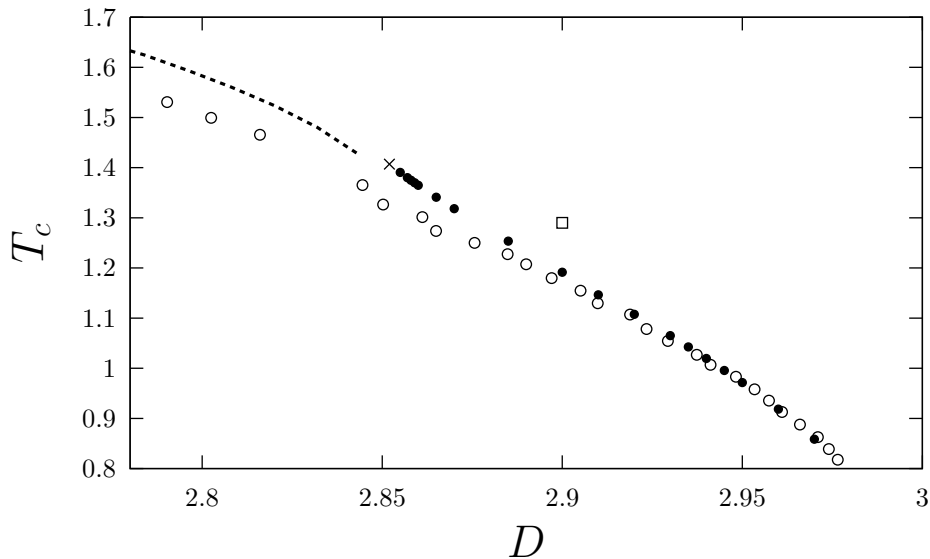}
\caption{\label{figure5}Zero-field FOPT temperatures (black points)
and the TCP ($\times$) at different values of anisotropy $D\geq D_{tr}$
calculated within SC-LPA RG approach. For comparison are shown the FOPT
and the continuous transition temperatures (open circles and the dashed
line) calculated from TSE in \cite{butera_blumecapel_2018}. The
square is a FOPT temperature obtained in numerical simulations in 
\cite{ozkan_critical_2006}. For further explanation see the text.}
\end{figure}

In the present paper all above methods have been used in the determination
of the transition temperatures shown in figure \ref{figure5}. At
many points two or three of them have been compared and within
the accuracy of calculations no discrepancies were found. Part
of the results of \cite{butera_blumecapel_2018} obtained via TSE
are also shown in figure \ref{figure5} for comparison. The authors
pointed out that the FOPTs temperatures obtained from the condition
of equality of the FE in the two phases seems to be systematically
underestimated. They arrived at this conclusion by application of the
method to the continuous transitions with $T_c$ reliably determined by
different technique (see figure \ref{figure5}). The RG points seems to
correct this deficiency, in particular, because they linearly approach
TCP which coordinates (\ref{DtTt}) have been determined with high
accuracy. The linear slope was practically the same as above TCP (for
the data shown in figure \ref{figure2}) which is the expected behaviour
\cite{butera_blumecapel_2018}. The transition temperature at the point
taken from the numerical simulations of \cite{ozkan_critical_2006}
seems to be too high to fall on the line passing through TCP with the
same slope as in the continuous transitions range. Thus, it is quite
plausible that the SC-LPA RG equation predicts FOPT temperatures in the
BCM more accurately than the currently available TSE and the numerical
simulation methods.
\subsection{Magnetization and the nonordering density at FOPTs}
\begin{figure}[htp]
	\centering \includegraphics{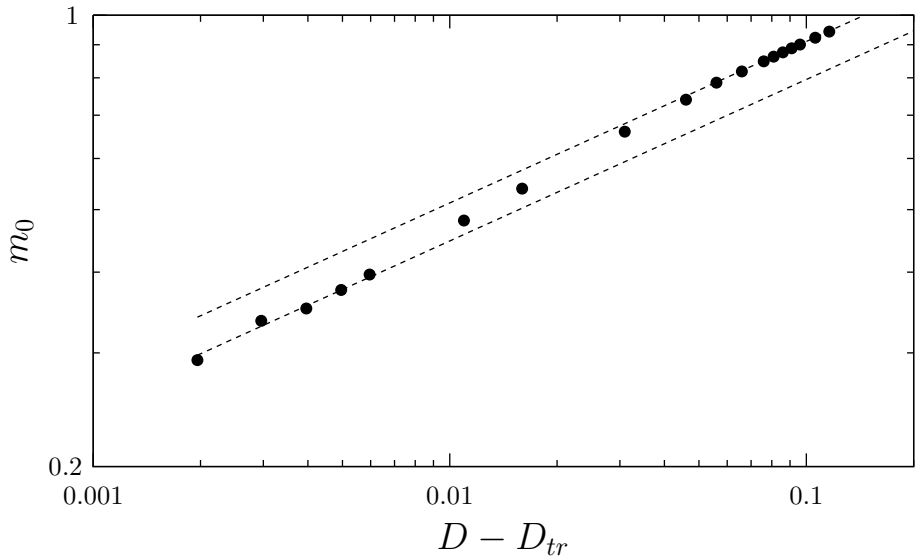}
	\caption{\label{figure6} Symbols---dependence of spontaneous
magnetization at FOPT on the distance to $D_{tr}$ calculated in the
SC-LPA RG approach; dashed lines---power-law dependence $(D-D_{tr})^{1/4}$
fitted to the data in the two regions discussed in the text.}
\end{figure}

Below $T_{tr}$ the spontaneous magnetization at FOPT temperature $T_c$
is finite but approaches zero as $T_c\to T_{tr}$; $m_0(T_c)$ dependence
calculated in the present approach is shown in figure \ref{figure6}.
As is seen, the dependence can be roughly separated into a low-temperature
region and a near-TCP region separated by a temperature interval with
transient behaviour.  In both regions similar power-law behaviour can
be observed but with different amplitudes.  Unfortunately, reliable
independent data on $m_0(T_c)$ are apparently absent in literature.
Understanding this unusual behaviour would be important for the present
study because the magnetization significantly influences the latent heat
which will be calculated below in the near-TCP region.  Of major concern
in this respect is the power-law behaviour with exponent 0.25 near TCP
which is very steep so the numerical solution of the RG equation with
the use of a relatively large step ($\Or(10^{-3})$) in $y\propto m$ could
introduce systematic errors into $m_0$. The step size was dictated by
the software used in the calculations \cite{tokar2019,sc-lpa-rg} so to
resolve the issue calculations within independent techniques and/or the
use of better precision software \cite{caillol_non-perturbative_2012}
would be needed.

\begin{figure}[htp]
\centering \includegraphics{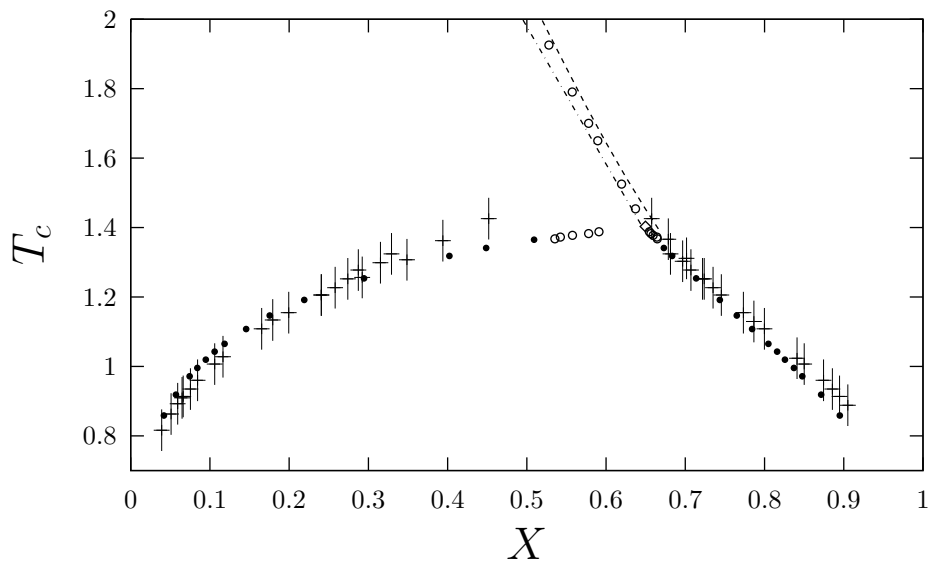}
\caption{\label{figure7} Dependence of the phase transition temperature
on non-ordering density $X=1-\langle s_i^2\rangle$. The dashed line and
the crosses are the TSE data from \cite{butera_blumecapel_2018};
the open circles and the diamond (at TCP) have been calculated within
SC-LPA approach according to (\ref{X1}); the black dots and the
dashed-dotted line have been calculated with the use of (\ref{X2})
with $G^R$ approximated by $G$.}
\end{figure}

The non-ordering density \cite{riedel_scaling_1972} is equal to the
concentration of sites with $s_i=0$. It can be found by subtracting the
ordering density, that is, the density of magnetic atoms with $s_i=\pm1$
from the total {\em per site} density which is equal to one
\begin{eqnarray}
	\label{X1}
	X &=& 1-\langle s_i^2\rangle=1-df/d\Delta\\
	\label{X2}
	&=&1-m_0^2-G^R_{ii}
\end{eqnarray}
where site $i$ can be chosen arbitrarily due to the homogeneity.
Equation (\ref{X1}) follows from (\ref{f}) differentiated by $\Delta$ and
(\ref{X2}) from (\ref{Z}) differentiated twice by $h_i$.  Though
equivalent in the exact theory, the two definitions have different
status in the LPA because in (\ref{X2}) enters the fully renormalized
correlation function $G^R$ which is not calculated in the SC-LPA because
this requires knowledge of the full momentum dependence of $G^R({\bf
k})$ while in LPA according to (\ref{G=G}) only $k\to0$ limit is
available. Still, this approximation is not very bad as is illustrated
by the dashed-dotted curve in figure \ref{figure7}.

But the best accuracy should be expected with the use of (\ref{X1}). The
values calculated with the direct use of this equation above $T_{tr}$
are shown in figure \ref{figure7} by open circles. Their accuracy is
confirmed by two facts. First, at TCP the calculated $X_{tr}$ differs from
the MC value of \cite{deng_red-bond_2004} by only 0.2\% while the TSE
value of \cite{butera_blumecapel_2018} on almost 3\%.  Second, at the
negative end of $D$ values BCM becomes the Ising model. But introducing
term $\Delta s_i^2$ into the Ising model the initial local potential
will acquire additional constant term---$\Delta$---which will remain in
the fully renormalized FE unaffected by renormalization because the LPA
RG equations contain only derivatives. Thus, in the limit $D\to-\infty$
$df/d\Delta=1$---the exact average value of $s_i^2$ in the spin-1/2
Ising model.

In the FOPTs region below $T_{tr}$ the direct application of 
(\ref{X1}) becomes more difficult because there are two branches on
one of which the spontaneous magnetization have to be self-consistently
determined which degrades the accuracy. This makes calculation of $f$
at different $D$ needed in (\ref{X1}) either less accurate or
more computationally demanding if higher finite differences with larger
number of points would be used.  Therefore, for $D$ close to the limiting
value $D=3$ where the dependence on $D$ becomes strong, the densities
$X_{d,o}$ have been calculated with the simpler (\ref{X2}) with
$G^R\approx G$ (see figure \ref{figure7}) while near TCP (\ref{X1})
was used indirectly as follows. To reduce the computational effort by
using already calculated data, the derivative of $f(T_c,D)$ over $D$
along the FOPTs line which we denote by $C$ has been calculated as
\begin{equation}
	\left.\frac{df}{dD}\right|_C = 
	\left.\frac{\partial f}{\partial T}\right|_{T=T_c}\frac{dT_c}{dD}
	+\frac{\partial f}{\partial D}.
	\label{df/dD}
\end{equation}
In view of the definition $\Delta = D/T$, the last term on the r.h.s.\ 
of (\ref{df/dD}) can be used to write the non-ordering density in  
(\ref{X1}) as
\begin{equation}
	X_{d,o} = 1-T_c\left(\left.\frac{df}{dD}\right|_C - 
	\left.\frac{\partial f_{d,o}}{\partial T}\right|_{T=T_c}\frac{dT_c}{dD}\right).`
	\label{Xdo}
\end{equation}
Because at $T_c$ $f_o=f_d$, among the terms on the r.h.s.\ of this
expression only the temperature derivative is phase-dependent. Many such
derivatives already have been calculated in the extrapolations needed
to determine $T_c$ from the condition $f_o=f_d=f$ which automatically
provided also the value of $f(T_c,D)$. Thus, only already known quantities
can be used in (\ref{Xdo}). The derivatives over $D$, however, can
be reliably calculated only when the values are sufficiently narrowly
spaced.  This was the case for several points closest to TCP where $\Delta
D=10^{-3}$ has been used. $X_{d,o}$ calculated in this way are shown in
figure \ref{figure7} below $T_{tr}$ by open circles. Unfortunately, the TSE
data are unavailable in this region but as can be seen the circles lie on
the line interpolating the black points calculated via (\ref{X1})
toward TCP. The black points agree well with the TSE data farther from
TCP but the agreement worsens on its approach. This divergence, however,
is most probably due to inaccuracy of the TSE calculations because the
TSE points closest to TCP have temperature coordinates larger than
$T_{tr}$ which is impossible because above the tricritical temperature
the transitions should be continuous, not of the first order. Thus, the RG
calculations agree with the TSE farther from TCP, correct the unphysical
behaviour of TSE on approach to TCP and, besides, can be carried out in
the most interesting region in the vicinity of TCP.
\subsection{The latent heat}
\begin{figure}[htp]
\centering \includegraphics{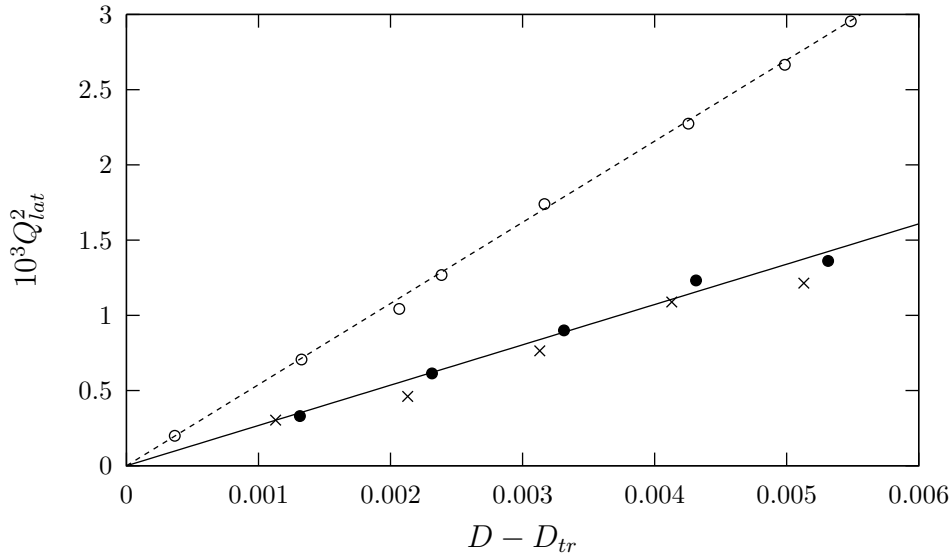}
\caption{\label{figure8}Latent heat for anisotropy $D$ in the vicinity
of TCP. Black points---calculations according to (\ref{Qlat})
within the present approach; open circles---the data taken from 
\cite{deserno_tricriticality_1997}. The lines are the best fits to
the data. For comparison are shown the points calculated 
according to (\ref{QlatX}) ($\times$).}
\end{figure}

Determination of the latent heat near TCP is one of the least studied 
problems in BCM because only one calculation by the MC
technique seems to be available \cite{deserno_tricriticality_1997}. Part
of the data \cite{deserno_tricriticality_1997} are shown in figure
\ref{figure8} together with the values calculated within the present
approach. In the RG calculations the difference between the internal
energies $e$ \cite{deserno_tricriticality_1997} in the disordered ($d$)
and ordered ($o$) phases 
\begin{equation}
	Q_{lat}=e_d-e_o
	\label{Qlat}
\end{equation}
was found with the use of the Gibbs-Helmholtz equation that follows from
(\ref{f}) (with $h=0$) valid in both phases
\begin{equation}
	e = -T^2\partial f/\partial T.
	\label{e}
\end{equation}
In (\ref{Qlat}) and (\ref{e}) use has been made of the temperature
derivatives found in the linear interpolation of the free energies to 
determine $T_c$.

As can be seen, the calculated absolute values of $Q_{lat}$ disagree with
MC data on about 40\% in stark contrast with previous comparisons. However,
to draw definite conclusions about the accuracy of any of the data shown
in figure \ref{figure8}, additional studies would be necessary for the
following reasons.

The data of \cite{deserno_tricriticality_1997} do not correspond to the
thermodynamic limit but are calculated in a finite system with the linear
box size $L=10$ lattice units. The interpolation to $L=\infty$ was not
attempted presumably because simulations with larger box sizes did not
show the expected large-$L$ behaviour \cite{deserno_tricriticality_1997}.
But there are reasons to believe that in the thermodynamic limit the
absolute value of the latent heat will be essentially reduced. In
the simulation of FOPTs in the 3-state Potts model on the sc lattice
which is physically similar to BCM \cite{potts_eq_bc} it was found
\cite{wilson_multilattice_1987} that in the system with $L=32$ the latent
heat is 12\% smaller than in the system with $L=16$. Assuming the leading
correction to be due to the surface term, that is, assuming the latent
heat at finite sizes to behave as $Q_{lat}(1+\delta/L)$ it is easy to
find that at $L=10$ the heat will be $\sim40\%$ larger than $Q_{lat}$.
But to confirm this estimate the interpolation within MC simulations of
BCM or the TSE calculation would be necessary.

The SC-LPA RG calculation of $Q_{lat}$ also is not devoid of
difficulties.  The problems arise because of the steep behaviour of
the magnetization near TCP due to $\beta=0.25$ which necessitate the
solution of additional self-consistency equation in the ordered phase
\cite{sc-lpa-rg,tokar2019}. This is further aggravated by the fact that
the quantity of interest ($Q_{lat}$) vanishes at TCP so in its vicinity
it is small but can be found only as a difference between two quantities
($e_d$ and $e_o$) that are finite at the transition. Thus, even their
calculation with a good precision does not guarantee that the difference
will be equally accurate.

A convincing test of the accuracy of the RG calculation of
the latent heat could have been obtained from TSE simulations of
\cite{butera_blumecapel_2018} if the calculations of $X(T)$ diagram below
$T_{tr}$ were continued closer to TCP to the region of the open circles in
figure \ref{figure7}. In this case $Q_{lat}$ could be expressed through
the non-ordering density as follows.  With the use of (\ref{Qlat}),
(\ref{e}) and (\ref{Xdo}) it is easy to derive expression for the latent
heat in terms of $X$ as
\begin{equation}
	Q_{lat}=T_c\left|\frac{dT_c}{dD}\right|^{-1}(X_d-X_o).
	\label{QlatX}
\end{equation}
In the vicinity of TCP $T_c(D)$ dependence is approximately linear so
the difference of $X$ values in (\ref{QlatX}) 
behaves as \cite{riedel_tricritical_1972,butera_blumecapel_2018}
\begin{equation}
	\Delta X\simeq A_1(T_{tr}-T_c)^{1/2}\simeq A_2(D-D_{tr})^{1/2} 
	\label{DX}
\end{equation}
where $A_{1,2}$ are some constants.  So the good agreement between
$\Delta X$ calculated by TSE and the RG (black points) seen on the phase
diagram figure \ref{figure7} could mean that closer to TCP (the open
circles) the agreement would be similarly good and so TSE would
confirm quantitatively the RG calculations of the latent heat. Unfortunately,
the range of validity of (\ref{DX}) is unknown, except that it should
hold asymptotically close to TCP. But from (\ref{X2}) one gets
\begin{equation}
	\Delta X=m_0^2+\left(G^R_{ii}\right)_o-\left(G^R_{ii}\right)_d.
	\label{DX2}
\end{equation}
It has been found that in the approximation $G^R_{ii}\approx G_{ii}$ the
last two terms on the r.h.s.\ practically cancel each other so $\Delta
X\approx m_0^2$.  As discussed earlier, it would be impossible to fit
$m_0^2$ to the dependence (\ref{DX}) with the same constant $A_2$
in the whole $D-D_{tr}$ region shown in figure \ref{figure6}. Thus,
unfortunately, the available TSE data cannot be interpolated to be
compared with the RG and with MC estimates in the vicinity of TCP. To
clarify this issue, improved simulations by at least one of the three 
methods---MC, TSE and the SC-LPA RG---would be desirable.
\section{Conclusion}
In this paper the SC-LPA RG approach developed in
\cite{sc-lpa-rg,tokar2019} where it was applied to the description of
critical behaviour in 3d lattice models has been extended to additionally
describe TCP and FOPTs that appear in the most complete version of the
Landau-Ginzburg model with polynomial interaction of the sixth order
($\phi^6$ model) \cite{landau_statistical_1980}, thus completing the
RG solution of the model.  The sc BCM has been chosen to validate the
approach because, on the one hand, its MF solution contains all salient
features of the $\phi^6$ model \cite{butera_blumecapel_2018}, on the
other hand, being a spin-lattice model, it has been extensively studied
by such reliable techniques as the exact MC simulations and the TSE
\cite{deng_red-bond_2004,ozkan_critical_2006,hasenbusch_finite_2010,%
fytas_universality_2013,butera_blumecapel_2018} so a wealth of
quantitative data are available for quantitative comparison.

It has been found that the SC-LPA RG approach describes the continuous
transitions in BCM in zero external field with the same accuracy
( $\leq0.6\%$ ) as in previous studies \cite{sc-lpa-rg,tokar2019};
in particular, the TCP parameters have been determined with errors
$\lesssim0.2\%$ and the classical tricritical exponents have been
reproduced exactly in the LPA RG.

The most important advancements in comparison with 
\cite{sc-lpa-rg,tokar2019} concern the description of FOPTs. In
the only systematic quantitative study of BCM FOPTs in 
\cite{butera_blumecapel_2018} carried out by TSE method the authors
found that at the level of truncation of TSE adopted by them the
FOPT temperatures were systematically underestimated. The authors
assessed the discrepancies quantitatively via comparison with reliably
determined critical temperatures.  In the present study it has been
shown that the temperatures calculated by SC-LPA RG technique are close
to those that would be obtained if the TSE predictions were corrected
on the discrepancies.  Also for the diagram the transition temperature
versus non-ordering density the RG data seems to be more accurate and
consistent than the TSE ones.  Thus, it seems that at present the SC-LPA
RG approach gives more accurate and complete description of FOPTs in 3d
lattice models of the Landau-Ginzburg type than is currently accessible
to alternative approaches. However, in the absence of internal criteria
on the equations accuracy, more extended TSE and/or more accurate MC data
would be desirable to confirm the above statement.

It is pertinent to point out that the main advantage of the proposed
RG approach is its computational simplicity. While MC simulations
in principle are capable of producing more reliable and accurate
results, the description of one critical point may require the use
of five Linux clusters and still some quantities can be determined
inaccurately \cite{ferrenberg2018,sc-lpa-rg}. Besides, the treatment
of pair interactions of long extent is a challenging task for both MC
and TSE methods. The SC-LPA RG equations, in contrast, can be easily
solved on virtually any computer irrespective of the extent of the pair
interactions and the complete phase diagram can be drawn with much less
effort than in the MC approach. The latter, however, may be needed to
validate the RG data by checking their accuracy at a few points.

The accurate quantitative description of FOPTs achieved in the
SC-LPA RG approach lends support to the underlying picture of the
RG flow. The latter has been described with resort to the GBE that
can be straightforwardly derived from the SC-LPA RG equation in the
scalar case. It has been shown that by analogy with the well-studied
example of the IRIM \cite{Brankov_1983,choquard_mean_2004}, FOPTs in
the short-range models should correspond to the discontinuous shock wave
solutions of GBE. It has been argued that such solutions that normally do
not appear in the viscous equation should be observable in the equations
originating from the LPA RG case when the viscosity and its derivative
turn to zero at the end of renormalization. Though a rigorous proof
of this statement has not been given, in numerical simulations it has
been satisfied within the accuracy of the calculations $\Or(10^{-6})$
which should be sufficient for an approximate theory.

The RG description of FOPTs in IRIM and in the short range
models had not revealed any qualitative difference between the
two cases which in particular means that unlike in the case of
critical phenomena, the Landau mean field theory is qualitatively
correct in FOPT case.  This may be helpful in clarifying the
controversial issue of the discontinuity fixed point discussed in
\cite{nienhuis_first-order_1975,zia_variety_1981,fisher_scaling_1982,%
privman1982,maxwell_construction}. In the Landau theory thermodynamic
system is considered to be of arbitrary dimension, spatial correlations
between the system constituents are absent and the scaling does not
take place \cite{landau_statistical_1980}.  Therefore, the use in the
description of FOPTs of the scaling relations, especially those that
include the dimensionality \cite{fisher_scaling_1982}, can be misleading.

To sum up, the SC-LPA RG equations due to their simplicity, numerical
accuracy and versatility may serve as a viable alternative to existing
approaches in the approximate description of phase transitions in 3d
lattice models with local interactions.
\ack
This research was supported by the National Academy of Sciences of Ukraine
under contract No. 22/20-H. I am indebted to Hugues Dreyss\'e for his
support and encouragement. I expresses my gratitude to Universit\'e de
Strasbourg and IPCMS for their hospitality and to Ren\'e Monnier for
interest in the work.
\providecommand{\newblock}{}

\end{document}